\documentclass[sigconf,nonacm]{acmart}
\usepackage{popets}

\usepackage{cleveref}
\newcommand{\cmark}{\checkmark}
\usepackage{siunitx}

\acmYear{YYYY}
\acmVolume{YYYY}
\acmNumber{X}
\acmDOI{XXXXXXX.XXXXXXX}
\acmISBN{}
\acmConference{Proceedings on Privacy Enhancing Technologies}
\begin{document}

\title[Reliability of the Multi-Input Heuristic in the Context of Law Enforcement]{How Reliable Is the Multi-Input Heuristic for Bitcoin Address Clustering in Law Enforcement Contexts?}

\author{Leopold M{\"u}ller}
\affiliation{%
  \institution{University of Bayreuth}
  \country{Germany}}
  \email{leopold.mueller@uni-bayreuth.de}

\author{Jana Elsner}
\affiliation{%
  \institution{University of Bayreuth}
  \country{Germany}}

\author{Thomas Niedermayer}
\affiliation{%
  \institution{Iknaio Cryptoasset Analytics}
  \country{Austria}}

\author{Bernhard Haslhofer}
\affiliation{%
  \institution{Complexity Science Hub}
  \country{Austria}}

\author{Thomas Goger}
\affiliation{%
  \institution{Bavarian Central Office for the Prosecution of Cybercrime}
  \country{Germany}}

\author{Niklas K{\"u}hl}
\affiliation{%
  \institution{University of Bayreuth}
  \country{Germany}}

\author{Christian R{\"u}ckert}
\affiliation{%
  \institution{University of Bayreuth}
  \country{Germany}}





\renewcommand{\shortauthors}{M{\"u}ller et al.}


\begin{abstract}

Address clustering is an important technique in blockchain forensics, widely employed by law enforcement to trace illicit crypto asset flows. The multi-input heuristic (MIH), which clusters addresses potentially associated with the same entity, is the most widely used. Yet, despite its broad adoption, the MIH has rarely been evaluated against reliable ground truth data.
We implement a reusable evaluation framework covering nine established metrics and apply it to ground truth address-to-entity mappings obtained directly from European crypto asset service providers under legally mandated reporting obligations.
When evaluation is restricted to reported addresses, the MIH appears strong at dataset level: we observe no mergers between reported services and recover same-service address pairs with recall $0.71$. However, this result is driven by one large service and ignores unlabeled addresses absorbed into full clusters. Metrics that assess the full clusters show substantially lower precision and recall ($0.36$ and $0.44$), meaning that services are often only partially recovered or embedded in larger clusters. Entity-level results further reveal near-complete failures for some services.
When MIH-based clusters are used to support criminal suspicion, as ground for preliminary seizure of crypto assets to secure their later forfeiture (U.S.)/ confiscation (Ger.), or as evidence in trial proceedings, prosecutors and judges must account for the heuristic's metric-dependent and entity-dependent reliability.

\end{abstract}


\maketitle


\section{Introduction}
\label{sec:introduction}

Clustering crypto asset addresses associated with the same real-world entity is a well-established deanonymization technique in blockchain forensics. The underlying intuition is straightforward: once a single address is linked to a known entity, this attribution is often implicitly extended to all addresses within the same cluster~\citep{frowis2020safeguarding}. Among clustering methods, the multi-input heuristic (MIH), which assumes that all input addresses of a transaction are controlled by the same entity, is the most widely adopted. It is routinely employed either as a standalone clustering rule or as a core component within larger heuristic pipelines~\citep{meiklejohn2013fistful,reid2012analysis,androulaki2013evaluating,ron2013quantitative,harrigan2016unreasonable,ermilov2017automatic,goldfeder2018cookie,he2022bitcoin,kappos2022peel,moser2022resurrecting,tubino2022towards,schnoering2024assessing,gong2025improved,zhang2020heuristic,chaudhari2021towards,lubbertsen2025ghost}. However, despite this widespread adoption in both research and practice, the heuristic has not been rigorously validated against recent real-world ground truth data that is independent of heuristic assumptions. 

In criminal proceedings around the world, clustering results can, for example, form a criminal suspicion against the owner of the address in question to request address information from cryptocurrency exchanges. They can also be used as the basis for a preliminary seizure to secure the subsequent judicial forfeiture (U.S.)/ confiscation (Ger.) of the seized crypto assets, or as additional evidence in court to prove the defendant's guilt. Although these results are usually only used as supplementary evidence and not as the sole basis for procedural measures, prosecutors and judges still need to know how reliable the underlying clustering method (such as the MIH) is for the following reasons:

In order to proceed further with the legal process (i.e., to be able to take certain investigative measures, file charges, or determine guilt at trial), prosecutors (and later judges) require legally defined levels of suspicion or (judicial) persuasion. All of these levels have two elements in common: a factual basis and a probability forecast that the suspect (later the defendant) has committed a (this) prosecutable crime. The degree of the criminal suspicion or persuasion required for both elements depends on the stage of the proceeding and the extent to which the following procedural measures affect the individual's fundamental rights. 
To generate certain levels of criminal suspicion or persuasion based on clustering results, the reliability of underlying clustering methods (such as the MIH) influences whether the results provide the required factual basis and also shapes the respective probability forecast. Thus, from a legal perspective, the reliability of the MIH is a decisive factor when the clustering results based on this method should (additionally) lead to certain procedural measures or judicial decisions. 

However, previous research has identified various metrics for the MIH that correspond to different performance levels. From a legal perspective, it is therefore particularly important to first define which of these metrics are relevant to prosecutors and judges and then (using a ground-truth dataset) to determine the values of the selected metrics.

To date, only three studies provide systematic MIH evaluations~\citep{remy2017tracking,nick2015data,gong2022analyzing}. However, their ground truth sources are either limited in scope and tied to outdated historical contexts~\citep{nick2015data,remy2017tracking}, or generated through simulation~\citep{gong2022analyzing}, each introducing assumptions that may not reflect contemporary blockchain usage. This limitation stems largely from the scarcity of reliable ground truth data~\citep{ermilov2017automatic}. Furthermore, existing evaluations report heterogeneous metrics, including pairwise scores, per-wallet scores, normalized mutual information (NMI), adjusted NMI (aNMI), and average error rate (AER). This heterogeneity yields a fragmented picture of MIH performance and leaves unclear how reported numbers should be interpreted, particularly in law enforcement settings where the costs of false positives and false negatives can be asymmetric.

The goal of this study is to evaluate the MIH against reliable ground truth data. To this end, we use pseudonymized address-to-entity mappings of European crypto asset service providers whose controlled blockchain addresses are subject to statutory reporting obligations.
To our knowledge, this constitutes a globally unique dataset for MIH evaluation. Building on this foundation, we pursue three research objectives. First, we investigate which evaluation methods have been applied to assess the performance of the MIH and how they differ in terms of procedures and output metrics (RO1). Second, we implement an evaluation framework and examine what performance state-of-the-art evaluation methods report for the MIH when applied to the same real-world ground truth data (RO2). Third, we outline the legal relevance for a reliability measurement of the MIH in three selected use cases where MIH-based clustering results are used as an investigative tool in criminal proceedings under German and U.S. criminal procedure law. (RO3). This paper makes three contributions:

\begin{enumerate}
    
    \item We unify existing MIH evaluations by reimplementing the preprocessing pipelines and all nine metrics reported across \citet{remy2017tracking,nick2015data,gong2022analyzing}. This enables direct comparison under consistent definitions. Our evaluation framework, including the implementations of all metrics and the preprocessing, is publicly available and can be applied to other datasets and clustering heuristics.\footnote{\url{https://anonymous.4open.science/r/mih-toolkit}}

    \item By applying our evaluation framework, we provide the first systematic evaluation of the MIH on verified ground truth data. We find that reported performance varies depending on the metric and aggregation level. We find a substantial gap between pairwise and per-wallet metrics: pairwise precision is perfect ($1.0$) with moderate recall ($0.71$) and F1 ($0.83$), per-wallet performance is substantially lower (precision $0.36$, recall $0.44$, F1 $0.27$), and global clustering agreement is limited (NMI $0.41$, aNMI $0.36$; AER $0.51$). Entity-level analysis further reveals heterogeneity across entities, and a leave-one-out sensitivity analysis shows that a single large entity can dominate dataset-level scores.
    
    \item We interpret our findings through the lens of US\ and German legal frameworks and discuss challenges when MIH-based clustering results are used as evidence, as grounds for suspicion, or as the basis for asset seizure.
    
\end{enumerate}

Overall, our findings show that MIH performance is highly sensitive to both the chosen metric family and the aggregation level (whether performance is measured on dataset-level as average over all entities or on entity-level for each entity separately) at which results are reported. Dataset-level scores can appear reassuring even when MIH fails almost completely for individual services, while other services may still be clustered very reliably. Pairwise metrics in particular can mask severe cluster contamination when evaluated only on labeled subsets. This has two direct implications. For research, it motivates more transparent and standardized reporting practices that include entity-level results (or distributions across entities) rather than dataset-level averages alone. For law enforcement practice, it reinforces that MIH-based clustering should be treated as an investigative lead rather than an absolutley reliable attribution mechanism. When clustering is used as the basis for criminal investigation measures, the service-dependent failure modes and the asymmetric risks of false positives must therefore be taken into account.


\section{Background \& Related Work}
\label{sec:background}

This section introduces the key terminology used throughout the paper (\cref{subsec:terminology}), situates our work within the broader landscape of blockchain forensics (\cref{subsec:forensics}), and reviews prior evaluations of the MIH, highlighting their data and metric choices and the resulting limitations for comparability (\cref{subsec:prior_evaluations}).

\subsection{Terminology}
\label{subsec:terminology}

In this paper, we focus on the crypto currency Bitcoin. Bitcoin \emph{transactions} transfer value between \emph{addresses}, which are pseudonymous identifiers derived from public keys. Each transaction consumes one or more unspent transaction outputs (UTXOs) as inputs and creates one or more new outputs, each assigned to an address. A \emph{wallet} is software or hardware that manages a collection of private keys and thereby controls the corresponding addresses. The term \emph{entity} refers to the real-world actor (whether an individual, organization, or exchange) that ultimately controls one or more wallets and their associated addresses.

In this paper, we use the term \emph{service} to denote entities that provide public-facing functionality on the blockchain (such as exchanges, payment processors, or gambling platforms); in our ground truth dataset, each service is represented by the set of addresses it controls. In regulatory contexts, many such services are classified as crypto-asset service providers (CASPs), i.e., organizations that professionally offer crypto-asset services such as custody, trading, or exchange.

\subsection{Blockchain Forensics}
\label{subsec:forensics}

Research in blockchain forensics covers a broad range of analytical tasks built on top of blockchain data. A first line of work uses network and graph analysis to derive structural patterns and cluster behavioral motifs, for example to characterize typical transaction behavior of specific entity types~\citep{reid2012analysis,ron2013quantitative,fleder2015bitcoin,tippe2025unmixing}. A second strand focuses on address-level classification, such as distinguishing change from non-change addresses or inferring the type of entity an address belongs to~\citep{toyoda2018multi,lin2019evaluation,tubino2022towards}. Moving from addresses to transactions, several studies classify transactions themselves (e.g., illicit vs.\ non-illicit or CoinJoin vs.\ non-CoinJoin)~\citep{weber2019anti,wu2021towards,schnoering2023heuristics,hu2019characterizing,hirshman2013unsupervised}. On top of this, a growing body of work seeks to classify entities or services (for example exchanges, gambling sites, or darknet markets) based on features aggregated over their on-chain activity~\citep{tovanich2023fingerprinting,bartoletti2018data,harlev2018breaking,jourdan2018characterizing}.

A key prerequisite for entity-level analysis is address clustering. Clustering is often used as a preprocessing step to derive entity-level features (such as the number of addresses controlled or transactions performed) that downstream classifiers operate on~\citep{harlev2018breaking}. Address clustering is therefore central not only for deanonymization, but also as an enabling component in many of the tasks above. The MIH is the most widely used clustering heuristic in this context, employed either standalone or in combination with other heuristics~\citep{meiklejohn2013fistful,androulaki2013evaluating,nick2015data,harrigan2016unreasonable,ermilov2017automatic,gong2022analyzing,he2022bitcoin,kappos2022peel,moser2022resurrecting,schnoering2024assessing,lubbertsen2025ghost,gong2025improved,zhang2020heuristic,remy2017tracking,goldfeder2018cookie}. Despite its central role, MIH is rarely evaluated in isolation.

From a legal and governance perspective, prior work highlights the risks of relying on heuristic-based analytics without robust validation and the need for systematic evaluation of crypto-forensics tools~\citep{frowis2020safeguarding,deuber2022sok}.~\citet{frowis2020safeguarding} propose a framework for more effective investigations that safeguards evidential value and fundamental rights, but empirical evaluations of concrete tools and heuristics remain scarce. Recent work by~\citet{lubbertsen2025ghost} evaluates the clustering performance of a specific commercial tool on three illicit services. They state that false positives were extremely rare for the examined tool. However, their evaluation treats a specific commercial tool as a black box without assessing the underlying clustering methods or their combination, which remain proprietary and undisclosed. This limits the ability to draw conclusions about the reliability of MIH itself.

This need for transparent validation is reinforced by~\citet{reiter2025out}, who shows that MIH can fail in scenarios that intentionally violate its core assumption. He proposes an out-of-sample stress test by transforming real Tornado Cash transactions on Ethereum into Bitcoin-style transactions with CoinJoin-like multi-party inputs. While this does not constitute Bitcoin ground truth, it illustrates the risk of large false-positive clusters under realistic multi-party spending behavior and motivates systematic validation of MIH on real-world Bitcoin ground truth.

\subsection{Prior MIH Evaluations}
\label{subsec:prior_evaluations}

To date, only three studies have systematically evaluated the standalone MIH~\citep{nick2015data,remy2017tracking,gong2022analyzing}. Each relies on distinct ground truth sources and evaluation metrics, which complicates comparison and underscores the need for a unified assessment on recent real-world data.

\citet{nick2015data} constructs a ground truth dataset of \num{37585} Bitcoin wallets by exploiting a vulnerability in BitcoinJ's BIP37 Bloom-filter implementation. Bloom-filter match-sets are used as approximations of wallet address sets, which are then cleaned and split into ``legacy'' and ``modern'' wallets depending on the wallet version and associated false-positive behavior. While this yields a comparatively large dataset, it is restricted to a specific simplified payment verification implementation, a short outdated time window, and a protocol vulnerability that has since been patched, limiting its relevance for today's broader and more diverse wallet ecosystem.

\citet{remy2017tracking} rely on a legacy public transaction dataset introduced by~\citet{meiklejohn2013fistful} comprising \num{16086073} transactions and \num{12056684} distinct addresses, and a manually curated ground truth of \num{776} labeled addresses grouped into \num{90} users obtained through interactions with various services. This labeled set is small, biased toward cooperative services, and confined to activity around 2013, so it may not reflect contemporary transaction patterns or the adoption of modern privacy techniques.

\citet{gong2022analyzing} use a different approach and evaluate MIH on synthetic data. They first extract empirical distributions from a full snapshot of the Bitcoin blockchain up to block height \num{687774} (16 June 2021), comprising roughly \num{649} million transactions, and then parameterize a transaction simulator that generates a small synthetic network of \num{300} nodes with complete, simulator-provided ground truth (\num{11796} transactions and \num{29005} addresses). This design enables controlled experiments but operates at limited scale and cannot fully capture the complexity, heterogeneity, and evolving behavior of real-world users.

Beyond the limitations of simulated, narrow, and outdated datasets, these three works also rely on different evaluation metrics.~\citet{remy2017tracking} report pairwise scores and information-theoretic measures (NMI, aNMI); ~\citet{nick2015data} focuses on per-wallet precision and recall; \citet{gong2022analyzing} report AERs. As summarized in~\cref{tab:related work}, no single study covers the full set of nine metrics, which complicates interpretation of MIH performance across settings.

\begin{table*}

  \caption{Comparison of evaluation metrics across prior MIH studies. Columns indicate whether each metric is reported (\cmark) or not reported (empty). Our study unifies all nine metrics under a consistent evaluation framework.}
  \label{tab:related work}
  \centering
  \resizebox{\linewidth}{!}{
  \begin{tabular}{l l ccc ccc ccc}
  \toprule
  & & \multicolumn{3}{c}{Pairwise} & \multicolumn{3}{c}{Per-Wallet} & \multicolumn{3}{c}{} \\
  \cmidrule(lr){3-5} \cmidrule(lr){6-8}
  Related Work                                     & Dataset                & Precision         & Recall          & F1           & Precision           & Recall & F1    & NMI    & aNMI   & AER\\
  \midrule
  \citet{remy2017tracking}                         & Real services (2013, small)   & \cmark            & \cmark          & \cmark        &                     &        &        & \cmark & \cmark &      \\
  \citet{gong2022analyzing}                        & Synthetic              &                   &                 &               &                     &        &        &        &        & \cmark \\
  \citet{nick2015data} (legacy wallets)            & Real wallets (2014--15)           &                   &                 &               & \cmark              & \cmark &        &        &        &      \\
  \citet{nick2015data} (modern wallets)            & Real wallets (2014--15, approx.)   &                   &                 &               &                     & \cmark &        &        &        &      \\
  \textbf{Our study}                                    & Real services (June 2023)           &  \cmark           & \cmark          & \cmark        & \cmark              & \cmark & \cmark & \cmark & \cmark & \cmark \\
  \bottomrule
\end{tabular}	
  }
\end{table*}

In summary, despite its widespread adoption, the MIH has not been rigorously validated against recent real-world ground truth that is independent of heuristic assumptions. Furthermore, heterogeneous metrics across existing evaluations yield a fragmented picture of MIH performance, leaving unclear how results should be interpreted in law enforcement settings. We address these gaps by reimplementing all nine metrics and applying them to ground truth obtained directly from crypto assets service providers under legal mandate, and discuss implications for both research and law enforcement practice.


\section{Data and Methods}
\label{sec:methodology}

This section presents our ground truth dataset (\cref{sub:ground_truth_dataset}), formalizes the MIH as used in this paper (\cref{sub:multi_input_heuristic}), and defines the cluster evaluation metrics applied in the remainder of this study (\cref{sub:cluster_evaluation_metrics}).

\subsection{Ground Truth Dataset}
\label{sub:ground_truth_dataset}

Our evaluation relies on pseudonymized address-to-entity mappings obtained from European crypto asset service providers under statutory reporting obligations. Under the applicable regulatory framework, these service providers are legally required to report blockchain addresses under their control to competent financial authorities. This provides a high degree of trustworthiness and completeness, distinguishing our dataset from prior work that relied on voluntary disclosures, simulation, or protocol vulnerabilities.

\begin{figure}[tb]
  \centering
  \includegraphics[width=\linewidth]{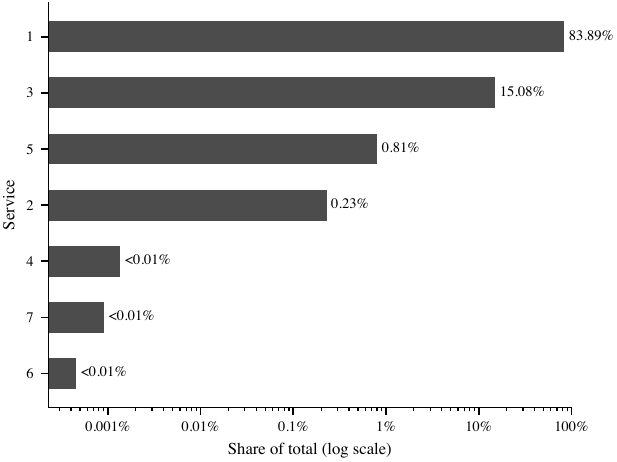}
  \caption{Relative size distribution of ground truth clusters for the seven services in our evaluation dataset. The distribution is highly skewed, with service~1 containing the majority of addresses. Services are pseudonymized and indexed from 1 to 7.}
  \label{fig:groundtruth_data}
  \Description{Relative size distribution of ground truth clusters for the seven services in our evaluation dataset. The distribution is highly skewed, with service~1 containing the majority of addresses. Services are pseudonymized and indexed from 1 to 7.}
\end{figure}

Specifically, the dataset contains pseudonymized Bitcoin addresses reported by seven services, with a cutoff date of 23 June 2023 (corresponding to Bitcoin block height \num{795357}). To protect both reporting entities and individual address holders, all service identifiers and addresses have been pseudonymized. Each service is represented solely by the set of addresses it controls without disclosure of actual identities. Let $R=\{R_1,\ldots,R_7\}$ denote these ground truth clusters.
The relative size distribution of these clusters is shown in \cref{fig:groundtruth_data}.
The support distribution is imbalanced. Service~1 dominates the labeled address set, service~3 has moderate support, services~2 and~5 have low support, and services~4,~6, and~7 are represented by very few addresses. Service~6 constitutes a degenerate singleton case, as it contains only one labeled address. Consequently, its metric values may distort dataset-level averages.

To construct the transaction corpus, we extracted all Bitcoin transactions from a full node up to block height \num{795357}, corresponding to the cutoff date. Let $\mathcal{T}$ denote this set of transactions. For each transaction $t \in \mathcal{T}$, let $I(t)$ denote the set of input addresses consumed by $t$. The induced address domain is $\mathcal{A} = \bigcup_{t \in \mathcal{T}} I(t)$, i.e., the set of all addresses that ever appear as transaction inputs.

In total, our corpus comprises $|\mathcal{T}| = \num{854144149}$ transactions with a mean of $\frac{1}{|\mathcal{T}|} \sum_{t \in \mathcal{T}} |I(t)| = 2.22$ inputs per transaction, covering $|\mathcal{A}| = \num{1105300685}$ distinct input addresses. Following prior work~\citep{nick2015data,remy2017tracking,gong2022analyzing}, we do not apply filters to remove potential CoinJoin or mixing transactions, and all subsequent computations use this unfiltered corpus. This keeps our setup directly comparable to the existing evaluations discussed in \cref{subsec:prior_evaluations}. Note that CoinJoin or mixing transactions violate the assumption of the MIH, and the picture may change when filters are applied.

\subsection{Multi-Input Heuristic}
\label{sub:multi_input_heuristic}

\begin{figure}[tb]
  \centering
  \includegraphics[width=\linewidth]{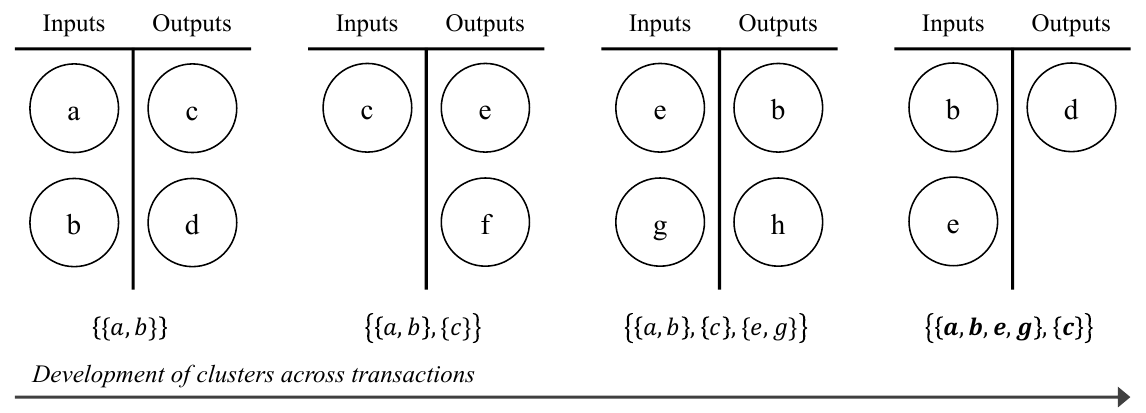}
  \caption{Illustration of cluster formation under the MIH. In each transaction, input addresses appear on the left and output addresses on the right. The set below each transaction shows the current partition of input addresses into clusters after processing transactions from left to right. Clusters merge whenever addresses appear together as inputs; transitive links can merge previously separate clusters (e.g., the final transaction with inputs $b$ and $e$ merges $\{a,b\}$ and $\{e,g\}$ into $\{a,b,e,g\}$).}
  \label{fig:mih}
  \Description{Illustration of cluster formation under the MIH. In each transaction, input addresses appear on the left and output addresses on the right. The set below each transaction shows the current partition of input addresses into clusters after processing transactions from left to right. Clusters merge whenever addresses appear together as inputs; transitive links can merge previously separate clusters (e.g., the final transaction with inputs $b$ and $e$ merges $\{a,b\}$ and $\{e,g\}$ into $\{a,b,e,g\}$).}
\end{figure}

The MIH rests on the assumption that all input addresses of a standard Bitcoin transaction are controlled by the same entity. Whenever two addresses appear together as inputs in a transaction, they are treated as belonging to the same entity. Repeating this reasoning transitively across chains of shared inputs yields the computed clusters $H=\{H_1,\dots,H_k\}$. \Cref{fig:mih} illustrates this process: addresses that appear together as inputs are merged, and subsequent transactions can merge previously separate clusters through transitivity.

\begin{figure}[tb]
  \centering
  \includegraphics[width=\linewidth]{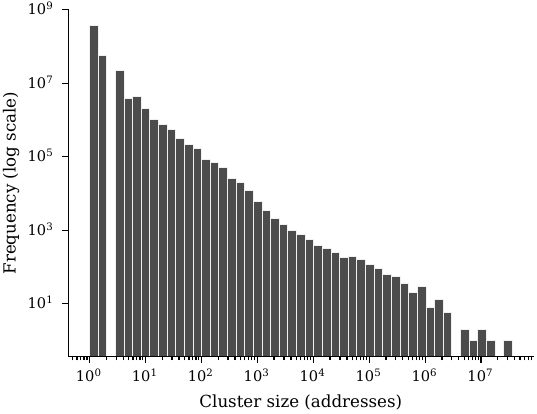}
  \caption{Distribution of computed cluster sizes (number of addresses per cluster) after applying the MIH to the transaction corpus.}
  \label{fig:found_clusters}
  \Description{Distribution of computed cluster sizes (number of addresses per cluster) after applying the MIH to the transaction corpus.}
\end{figure}

In our setting, the MIH is applied to the full transaction corpus $\mathcal{T}$ described above. For each transaction $t\in\mathcal{T}$, we consider its set of input addresses $I(t)$. Two addresses are linked if they appear together in at least one input set $I(t)$. The MIH then repeatedly merges all addresses that are directly or transitively connected in this way. The result is a partition of the observed input-address domain $\mathcal{A}$ into disjoint clusters $H_1,\dots,H_k$, where each cluster is interpreted as the set of addresses that the heuristic attributes to one entity.
\Cref{fig:found_clusters} shows the size distribution of the \num{466792735} computed clusters. The distribution is skewed, with most clusters containing few addresses and a small number of clusters aggregating large portions of the address space.

\subsection{Cluster Evaluation Metrics}
\label{sub:cluster_evaluation_metrics}

We evaluate the MIH by comparing the reported ground truth entities with the clusters produced from the full transaction corpus. Because labels are available only for the addresses reported by the seven services, the evaluation uses two complementary views. Metrics that require a known ground truth relation for every compared address, such as pairwise precision, pairwise recall, pairwise F1-score, NMI, and aNMI, are computed on the labeled subset only. Metrics that describe the clusters reached from labeled entities or the recovery of those entities, such as per-wallet precision, per-wallet recall, per-wallet F1-score, and AER, relate the labeled entities to the full MIH clusters.
This metric selection combines measures used in prior evaluations of address clustering~\citep{remy2017tracking,gong2022analyzing,nick2015data}. We add the per-wallet F1-score (see \cref{tab:related work}). In addition, we report AER and the per-wallet metrics both at dataset level and at entity level. This is important in our setting because the ground truth entities are highly imbalanced, so dataset-level means alone can hide entity-specific behavior.
For the formal definitions below, let $R=\{R_1,\ldots,R_r\}$ denote the ground truth entities and $H=\{H_1,\ldots,H_k\}$ the computed clusters produced by the MIH. We write $S=\bigcup_{i=1}^{r} R_i$ for the set of all labeled addresses and, when needed, restrict the computed clustering to this domain by intersecting each computed cluster with $S$, yielding $H^{S}=\{H_j\cap S \mid H_j\in H,\ H_j\cap S\neq\varnothing\}$.

\paragraph{Pairwise precision, recall, and F1-score.}
Pairwise metrics treat clustering as a binary decision over unordered pairs of addresses: two addresses are either placed in the same cluster or in different clusters. Intuitively, this evaluates whether the MIH makes the correct link / no-link decision for labeled address pairs. Pairwise precision measures how often address pairs that the MIH places in the same cluster truly belong to the same ground truth entity. Pairwise recall measures how many of the true within-entity address pairs are recovered by the MIH. The pairwise F1-score summarizes this trade-off as the harmonic mean of precision and recall.
Because ground truth labels are available only for a subset of addresses, we compute the pairwise metrics on the labeled domain $S=\bigcup_{i=1}^{r} R_i$ and compare the ground truth partition $R$ with the restricted computed partition $H^{S}$. This follows~\citet{remy2017tracking} and ensures that every counted pair has a well-defined ground truth relation. In other words, unlabeled addresses do not contribute to the pairwise counts. The required true-positive, false-positive, and false-negative counts are obtained from the overlap structure between $R$ and $H^{S}$, so pairwise metrics evaluate only labeled address pairs and do not reflect contamination by unlabeled addresses in the full MIH clusters. The resulting precision, recall, and F1-score are then computed in the standard way.

\paragraph{Per-wallet precision, recall and F1-score.}

Following~\citet{nick2015data}, we additionally evaluate clustering quality with per-wallet metrics. In this terminology, a wallet corresponds to one ground truth entity in our pseudonymized service data. For each labeled seed address $a\in R_i$, let $H(a)$ be the computed MIH cluster containing $a$. Per-wallet precision asks what fraction of the other addresses in $H(a)$ are also labeled as belonging to $R_i$, while per-wallet recall asks what fraction of the other labeled addresses in $R_i$ are recovered in $H(a)$. Thus, the seed address itself does not make a cluster look correct by construction. Only additional recovered addresses contribute to the score.

In contrast to the pairwise metrics, which are computed only on the labeled domain $S$ and evaluate link/ no-link decisions for pairs of labeled addresses, the per-wallet metrics evaluate the full computed clusters $H_j$, including unlabeled addresses. As a result, per-wallet precision can be substantially lower than pairwise precision when MIH clusters absorb many unlabeled addresses.
The per-wallet recall also differs conceptually from pairwise recall. Pairwise recall measures how many true same-entity address pairs are recovered globally on the labeled domain, whereas per-wallet recall measures, from the perspective of a single starting address, how much of that entity’s address set is recovered in the cluster it reaches. Per-wallet recall is therefore more sensitive to fragmentation of an entity across multiple clusters and, after macro-averaging, gives each entity equal weight instead of being dominated by large entities with many address pairs.

Intuitively, per-wallet precision measures how pure the computed cluster is when an investigator starts from one address of an entity. It captures the typical extent to which reached clusters are contaminated with unrelated addresses.
Per-wallet recall measures how much of an entity's known address set is recovered when an investigator starts from one address of that entity. Depending on the entity, it can be entirely normal that recall is below one, since not every address belonging to the same entity is necessarily connected through co-spending behavior. In such cases, the heuristic cannot be expected to recover the full entity cluster. The per-wallet F1-score summarizes the trade-off between these two quantities as the harmonic mean of precision and recall at the entity level. For dataset-level reporting, we macro-average the entity-level per-wallet precision, recall, and F1-scores separately. Thus, the dataset-level per-wallet F1-score is the average of the entity-level F1-scores, not the harmonic mean of the dataset-level precision and recall values.

\paragraph{Normalized mutual information and adjusted normalized mutual information score.}
We evaluate agreement between the ground truth partition and the MIH clustering using information-theoretic scores. Intuitively, NMI measures how much the cluster assignment under one partition reduces uncertainty about the assignment under the other partition, capturing overall alignment rather than performance for a specific entity~\citep{lancichinetti2009detecting,vinh2009information}. The adjusted variant (aNMI) additionally corrects for agreement expected by chance, following the adjustment proposed by~\citet{vinh2009information} (which is also the definition used by~\citet{remy2017tracking}).
As in the pairwise setting, we restrict the evaluation to the labeled domain. We then compute NMI and aNMI between the ground truth cluster  and the restricted computed cluster. Concretely, we report the geometrically normalized NMI and its chance-corrected counterpart aNMI.

\paragraph{Average error rate.}
The AER as used by \citet{gong2022analyzing} summarizes, in an asymmetric and one-to-best fashion, how well each ground truth entity is captured by the computed clustering. Intuitively, AER asks the following question: for a given ground truth entity, which single MIH cluster matches it best, and what fraction of the entity’s addresses does this best-matching cluster recover? The per-entity error is then one minus this recovery fraction, and AER is the average of these per-entity errors over all entities.
In contrast to the pairwise and per-wallet metrics, AER does not evaluate all overlaps between the ground truth and the computed clustering. Instead, it considers only the single best-matching cluster for each entity.
As a result, AER primarily reflects coverage, i.e., how much of an entity is captured at all, and is comparatively insensitive to contamination by additional foreign addresses in that cluster (false positives). In addition to the dataset-level AER, we also report the corresponding entity-level values to expose heterogeneity across services.


\section{Results}
\label{sec:results}

In the following sections, we present the performance of the MIH measured in nine different metrics.

\subsection{Performance on Dataset-Level}
\label{subsec:results-dataset-level}

\Cref{tab:dataset_level} summarizes the performance of the MIH on our dataset alongside results reported in prior work. Overall, the pairwise scores we obtain are close to the literature: our pairwise F1-score ($0.83$) is comparable to \citet{remy2017tracking} (0.86), with a slightly lower pairwise recall ($0.71$ vs.\ $0.77$) but higher pairwise precision ($1.00$ vs.\ $0.98$). This indicates that, on the labeled domain $S$ used for pairwise evaluation, the heuristic very rarely merges addresses from different ground truth entities (high pairwise precision), while still recovering a substantial fraction of true within-entity pairs (moderate pairwise recall). This is intuitive as we do not expect the services to co-spend with each other. We emphasize that pairwise metrics are computed after restricting both partitions to $S$. As a consequence, high precision reflects the absence of cross-entity mergers among labeled addresses, while mergers into large clusters via unlabeled addresses are not penalized by this metric.

\begin{table*}[tbh]
    \caption{Dataset-level performance of the MIH in prior work and in this study. For the first four rows, we list the values exactly as reported by the original authors on their own datasets \citep{remy2017tracking,nick2015data,gong2022analyzing}. Empty cells indicate metrics that were not reported in the corresponding work. The last row (\textbf{Our study}) shows the performance of our unified reimplementation of all nine metrics applied to our recent real-world ground truth dataset of seven services. The per-wallet dataset-level values are macro-averages over entity-level scores; in particular, the per-wallet F1-score is the average of entity-level F1-scores, not the harmonic mean of the dataset-level precision and recall.}
    \label{tab:dataset_level}
    \centering
      \begin{tabular}{l ccc ccc ccc}
        \toprule
        & \multicolumn{3}{c}{Pairwise} & \multicolumn{3}{c}{Per-Wallet} & \multicolumn{3}{c}{} \\
        \cmidrule(lr){2-4} \cmidrule(lr){5-7}
        Related Work                                     & Precision         & Recall          & F1            & Precision           & Recall            & F1 & NMI  & aNMI & AER  \\
        \midrule
        \citet{remy2017tracking}                         & 0.98              & 0.77            & 0.86          &                     &                   &          & 0.89 & 0.65 &      \\
        \citet{gong2022analyzing}                        &                   &                 &               &                     &                   &          &      &      & 0.63 \\
        \citet{nick2015data} (legacy wallets)            &                   &                 &               & 0.91                & 0.80              &          &      &      &      \\
        \citet{nick2015data} (modern wallets)            &                   &                 &               &                     & 0.69              &          &      &      &      \\
        \textbf{Our study}                                    & 1.00               & 0.71            & 0.83          & 0.36                & 0.44             & 0.27     & 0.41 & 0.36 & 0.51 \\
        \bottomrule
      \end{tabular}
\end{table*}

In contrast, the per-wallet scores are markedly lower than those reported for earlier time periods. Relative to \citet{nick2015data}, per-wallet precision and recall decline from legacy wallets ($0.91$/$0.80$) to modern wallets (recall $0.69$) and further to our setting ($0.36$/$0.44$). The dataset-level per-wallet F1-score ($0.27$) is the macro-average of the entity-level F1-scores reported in \cref{tab:enity_level}, and therefore does not equal the harmonic mean of the dataset-level precision and recall values. Two factors likely contribute. First, our observation window spans more than a decade, during which wallet behavior and clusterability evolved (e.g., increased unspent transaction output (UTXO) management across accounts, consolidation practices, and wider use of privacy-enhancing features). This could lead to a higher number of FP. Second, per-wallet precision excludes the seed itself, penalizing clusters that fail to recover additional same-entity addresses. These design choices make the per-wallet calculation more sensitive to partial discovery than the pairwise calculation.

Information-theoretic measures show a similar pattern. Both NMI ($0.41$) and aNMI ($0.36$) are lower than the values reported by \citet{remy2017tracking} ($0.89$ and $0.65$, respectively). Because NMI/aNMI summarize alignment of the partitions’ co-membership structure (including cluster-size distributions), this drop is consistent with the per-wallet picture: the clusters predicted by the MIH align less tightly with ground truth.

Finally, the AER is smaller in our data ($0.51$) than in \citet{gong2022analyzing} ($0.63$). AER is asymmetric and ``one-to-best'': it rewards the fraction of each ground truth entity captured by its best-matching computed cluster, without penalizing extraneous addresses in that computed cluster. Hence AER can improve even when per-wallet precision degrades, provided that the best cluster for each entity covers a nontrivial share of that entity. Interpreting AER together with per-wallet precision/recall gives a more nuanced view: MIH still retrieves sizable cores of many entities (lower AER), but does so with substantial over- or under-coverage when the clusters are not restricted to the labelled set (low per-wallet precision and recall).

Taken together, the dataset-level results paint a mixed picture that depends strongly on how unlabeled addresses enter the evaluation: pairwise scores on the labeled domain suggest that MIH rarely merges labeled entities, while the lower per-wallet scores indicate that, in the full clusters, services are often either only partially recovered or embedded in large clusters that include many additional (unlabeled) addresses.

\subsection{Performance on Entity-Level}
\label{subsec:results-entity-level}

While dataset-level metrics provide a compact summary, they conceal substantial heterogeneity across services. \Cref{tab:enity_level} reports per-service per-wallet precision, recall, F1-score, and AER. Overall, the MIH performs well for one large service but exhibits pronounced failures for several others, with performance varying by orders of magnitude across entities.

\begin{table}[tbh]
  \caption{Per-service performance of the MIH (entity-level). Each row corresponds to one ground truth service $R_i$ and reports its support (number of labeled addresses), per-wallet precision, recall, and F1-score, as well as the AER with respect to the best-matching computed cluster.}
  \label{tab:enity_level}
  \centering
  \resizebox{\columnwidth}{!}{%
  \begin{tabular}{r c ccc c}
    \toprule
    & \multicolumn{1}{c}{} & \multicolumn{3}{c}{Per-Wallet} & \\
    \cmidrule(lr){3-5}
    Service
      & Support
      & Precision
      & Recall
      & F1
      & AER \\
    \midrule
    1 & high     & 0.950 & 0.730 & 0.825 & 0.153 \\
    2 & low      & 0.279 & 0.004 & 0.008 & 0.978 \\
    3 & moderate & 0.032 & 0.000 & 0.000 & 0.999 \\
    4 & very low & 0.000 & 0.333 & 0.000 & 0.667 \\
    5 & low & 0.256 & 0.042 & 0.072 & 0.796 \\
    6 & very low & 0.000 & 1.000 & 0.000 & 0.000 \\
    7 & very low & 1.000 & 1.000 & 1.000 & 0.000 \\
    \bottomrule
  \end{tabular}
  }
\end{table}


Service~1 dominates the ground truth in terms of support (high) and shows comparatively strong performance: per-wallet precision is high ($0.950$) and recall is substantial ($0.730$), yielding an F1-score of $0.825$ and a relatively low AER of $0.153$. For this service, the MIH recovers a large fraction of addresses while keeping the best-matching cluster comparatively clean. Given its size, service~1 also has a strong influence on any dataset-level averages.

In contrast, services~2 and~3 illustrate near-complete failure modes. Service~2 (support low) has low recall ($0.004$) and a correspondingly low F1-score ($0.008$), together with a very high AER ($0.978$), implying that even the best-matching MIH cluster captures only a negligible fraction of the service. Service~3, despite being the second-largest entity in the ground truth (support moderate), achieves essentially zero recall and F1 and an AER close to one ($0.999$), meaning that MIH fails to recover the entity beyond a vanishingly small overlap. Service~5 (support low) shows intermediate behavior: precision is moderate ($0.256$), but recall remains low ($0.042$), resulting in a low F1-score ($0.072$) and a high AER ($0.796$). This suggests that the MIH identifies only a small subset of the service’s addresses and that this subset is not large enough to yield meaningful reconstruction at the entity-level.

Finally, the smallest services (support very low) highlight edge cases. For services with very few labeled addresses, recall can be comparatively high even when precision is low, because capturing a single address already accounts for a large share of the entity. Service~4 (support very low) attains recall $0.333$ but precision $0.000$, resulting in F1 $0.000$ and AER $0.667$. Service~6 (support very low) has recall $1.000$ but precision $0.000$, which yields F1 $0.000$ while AER is $0.000$. This combination reflects an edge case (see \Cref{sub:ground_truth_dataset}) of the evaluation: because AER and recall primarily measure whether the labeled service addresses are covered, they can be perfect for a very small service, while per-wallet precision and F1 remain zero when the reached cluster does not recover additional same-service addresses beyond the starting point. Service~7 (support very low) achieves perfect scores across all metrics (precision/recall/F1 all $1.000$, AER $0.000$), illustrating that MIH can align perfectly with ground truth in small, favorable cases.

Taken together, the entity-level results show that MIH performance is highly entity-dependent: strong reconstruction for a dominant service co-exists with near-zero recovery for others, including one moderate-sized service (service~3). This further reinforces that entity-level reporting is essential for interpreting clustering quality and that dataset-level metrics alone can mask severe failures on specific investigative targets.

\subsection{Sensitivity of Pairwise Metrics to Service Composition}
\label{subsec:results-pairwise-sensitivity}
Pairwise precision, recall, and F1-score are defined globally over all labeled address pairs in the evaluation domain $S$. In particular, while within-service pairs contribute to true positives and false negatives, false positives arise from cross-service pairs and therefore cannot be attributed to one ground-truth service in a straightforward way.
To better understand how much individual services contribute to the dataset-level pairwise scores, we recompute the pairwise precision, recall, and F1-score in a leave-one-out fashion. For each service in turn, we remove all of its ground truth addresses from the labeled domain $S$ (and thus from the restricted computed partition $H^{S}$) and then recompute the pairwise metrics on the remaining services. The case ``None'' corresponds to the full labeled set $S$ with all seven services included. \Cref{tab:pairwise_leave_one_out} reports the resulting scores. Overall precision remains $1.000$ in all settings, reflecting that the MIH almost never merges labeled addresses from different services. In contrast, recall and F1-score are sensitive to the inclusion or exclusion of specific services.

\begin{table}[tbh]
  \caption{Leave-one-out sensitivity of pairwise metrics. Each row shows the pairwise precision, recall, and F1-score when all addresses of the indicated service are removed from the labeled domain $S$ before computing the contingency table and pairwise scores. ``None'' corresponds to using all seven services.}
  \label{tab:pairwise_leave_one_out}
  \centering
  \begin{tabular}{r ccc}
    \toprule
     & \multicolumn{3}{c}{Pairwise} \\
    \cmidrule(lr){2-4}
    Excl. service & Precision & Recall & F1 \\
    \midrule
    None   & 1.000 & 0.707 & 0.828 \\
    1      & 1.000 & \textbf{0.000} & \textbf{0.000} \\
    2      & 1.000 & 0.707 & 0.828 \\
    3      & 1.000 & \textbf{0.729} & \textbf{0.844} \\
    4      & 1.000 & 0.707 & 0.828 \\
    5      & 1.000 & 0.707 & 0.828 \\
    6      & 1.000 & 0.707 & 0.828 \\
    7      & 1.000 & 0.707 & 0.828 \\
    \bottomrule
  \end{tabular}

\end{table}

Excluding service~1 causes pairwise recall to collapse from $0.707$ to $0.000$, and the F1-score drops from $0.828$ to $0.000$. This shows that essentially all correctly recovered same-entity pairs in the labeled domain originate from service~1. Once this service is removed, MIH recovers virtually no within-service pairs for the remaining entities. Conversely, excluding service~3 slightly increases recall (from $0.707$ to $0.729$) and F1-score (from $0.828$ to $0.844$), indicating that service~3 contributes disproportionately many false negatives in the pairwise evaluation. Removing any of the other services has no measurable effect on the reported pairwise scores.

These findings confirm that dataset-level pairwise metrics can be dominated by a single high-support service, and that the presence of poorly clustered entities can depress global recall via false negatives. Reporting only aggregated pairwise scores can therefore mask extreme heterogeneity across services, reinforcing the need to complement dataset-level reporting with entity-level performance measures.


\section{Discussion}
\label{sec:discussion}

This paper provides the first systematic evaluation of the MIH on a real-world ground truth of seven Bitcoin services, using a unified implementation of nine metrics taken from prior work. At the dataset-level, MIH attains pairwise scores comparable to earlier studies, with very high pairwise precision and moderate recall, but substantially lower information-theoretic agreement (NMI, aNMI) and per-wallet performance than previously reported. At the entity-level, performance is highly heterogeneous: for some services MIH recovers large, relatively clean cores, while for others it almost completely fails to reconstruct the ground truth entity. Together, these results show that the perceived quality of MIH clustering depends strongly on the chosen metric and aggregation level, and that treating MIH output as ground truth is not warranted.

\subsection{Implications for Research}

\paragraph{Legal perspective on metrics.}
From a legal perspective, the underlying notions of true/false positives and true/false negatives play different roles. True positives (correctly clustered addresses of the same service) support investigative leads. False negatives (same-service addresses placed in different clusters or left unclustered) correspond to missed links and may reduce investigative efficiency. Since they leave offenders unidentified, they are particularly critical, as victims of the affected crimes may be left to bear the resulting damages. False positives (addresses incorrectly grouped with a service) are also critical: they risk attributing illicit activity to unrelated users and thus raise concerns about proportionality and due process~\citep{frowis2020safeguarding,deuber2022sok}. True negatives (correctly separated addresses) are numerically dominant in large graphs but largely irrelevant for decision-making.

Against this background, different metrics emphasize different legal risk profiles. Pairwise precision and recall summarize how well the method gets co-membership decisions right for address pairs on the labeled domain only. However, pairwise precision can remain high even when large clusters contain many unrelated addresses, as long as labeled cross-entity pairs are rare. Per-wallet precision and recall, in contrast, operate at the entity-level: per-wallet precision indicates how ``contaminated'' the cluster around a service is (false positives), whereas per-wallet recall reflects how much of the entity is recovered (false negatives). NMI and aNMI capture global alignment of the partitions, but are difficult to interpret in operational or legal terms. AER, finally, is optimistic with respect to over-coverage because it rewards how much of each entity is covered by its best cluster without penalizing additional addresses in that cluster.
We therefore argue for the development and standardization of a law enforcement-oriented clustering metric that more directly reflects the evidentiary and proportionality requirements for justifying investigative measures based on MIH outputs, as existing metrics only partially capture these needs.

\paragraph{Dataset-level and entity-level.}

Our results highlight that dataset-level averages can be deeply misleading when used in isolation. The dataset-level metrics suggest a mixed but moderately capable heuristic on the labeled domain. However, the per-service breakdown reveals that these averages are dominated by a small subset of large services for which MIH works reasonably well, while it fails almost completely for others. At the same time, very low-support services can produce edge cases in which individual metrics appear extreme because only very few labeled addresses determine the score. The leave-one-out sensitivity analysis further quantifies this and shows that a single large, well-clustered service can mask poor performance elsewhere, and that poorly clustered services can depress global scores.

For research, this implies that future evaluations of clustering heuristics should routinely report entity-level distributions (e.g., per-entity precision/recall/AER, or at least summary statistics across entities), in addition to dataset-level scores. For practice, especially in an investigative context, it suggests that MIH output should not be treated as uniformly reliable across services. Instead, analysts should be made aware that the quality of clusters can vary dramatically between entities, and that high dataset-level scores do not guarantee acceptable performance for any particular target. This reinforces calls from the legal literature to treat heuristic-based analytics as investigative leads rather than definitive evidence, and to document their limitations transparently~\citep{frowis2020safeguarding,deuber2022sok}.

\paragraph{Generalizability.}

Our study focuses on seven services for which high-quality ground truth is available. Even within this relatively homogeneous group of service-type entities, performance varies. This variability is likely to be even greater for other kinds of entities (e.g., individuals, small businesses, or sophisticated entities employing privacy-enhancing techniques). Moreover, for comparability of the results, we deliberately adhere to the preprocessing of earlier work and do not filter out CoinJoin or mixing-like transactions. In contemporary Bitcoin usage, such patterns, as well as increasingly complex UTXO management and consolidation strategies, are common. As a result, the MIH may perform differently (and often worse) for entities that make heavy use of these practices than for the services in our dataset. The key implication for generalizability is that even if MIH appears ``good enough'' for some well-behaved services, this does not justify assuming similar performance in other settings or time periods. Evaluations should be repeated as usage patterns evolve and extended to a broader range of entities. Practitioners should avoid extrapolating from limited ground truth to the full address space, and instead treat the kind of evaluation we present here as a lower bound on the validation needed.

\subsection{Implications for Law Enforcement}

In the field of cryptocurrency, where crime does not stop at national borders, an international perspective on the requirements of different legal systems, whether adversarial or inquisitorial, is essential. Therefore, we compare the US criminal procedure, which represents the Anglo-American adversarial system, with the German criminal procedure, which is an example of the continental European inquisitorial system. Although there are differences in legal reasoning, prosecutors and judges must consider reliability criteria in order to establish a certain degree of suspicion or (judicial) persuasion in both. 

All levels of legal suspicion or persuasion described below have two things in common. First, they require a factual basis. Second, based on this factual basis, a probability forecast must be created that either the suspect committed the prosecuted offense or the discovered assets originated from a criminal offense committed (by any suspect). From the perspective of prosecutors and (later) judges, the reliability of the underlying clustering method is directly relevant for establishing the required factual basis and respective probability forecast  \citep{frowis2020safeguarding}.  

This section examines three use cases in which clustering results are relevant in criminal proceedings, highlighting the underlying impact of the clustering method in each context. The following considerations apply to all possible clustering methods. However, for the legal-technical proposals, we will focus specifically on the MIH, because we only measured the performance for this heuristic. 

\subsubsection{First use case: Request for information from cryptocurrency exchanges}

When prosecutors discover a possible committed offence that was processed via blockchain, they use clustering to narrow suspicion to a particular address. Address clustering makes possibly incriminated addresses visible and therefore provides an initial approach to identify the perpetrator or the participant. 

Under German law, the request for address information from centralized cryptocurrency exchanges~\citep{pelker2021using,korver2019attribution,dyson2019challenges,frowis2020safeguarding} is an intervention pursuant to the German Code of Criminal Procedure (GER-CCP), Section 161 and Section 163 and therefore requires legal justification by a ``simple''~\citep{frowis2020safeguarding}  or ``primary'' suspicion against the holder of the address in question, according to (GER-CCP), Section 152 (2). This degree of suspicion requires specific indications (factual basis) that a suspect (the holder of the address) has possibly committed a prosecutable offence (low probability forecast). 
Prosecutors usually form this degree of suspicion based on a combination of address clustering results and additional information outside the blockchain \citep{emehelu2018shot}, such as correspondence with the victim, bank account details, or suspicious behavior on social media. 
Thus, the reliability of the underlying clustering method (such as the MIH) influences whether clustering results can provide additional clues that the address holder is the possible perpetrator or participant of the offense being prosecuted. Its reliability also influences whether the probability forecast can be made. However, as the threshold for establishing this level of suspicion is low and further information is usually available, the reliability of clustering methods is less important in this procedural state overall.

Under US law, address information can often already be obtained under the low requirements of a (grand jury) subpoena \citep{pelker2021using,korver2019attribution,emehelu2018shot,huang2015reaching,deuber2022sok}, because according to the ``third-party doctrine'', no one has a legitimate interest in the confidentiality of information voluntarily disclosed to third parties [United States v. Gratkowski, 2020]. Because requests to cryptocurrency exchanges can be obtained through a subpoena, clustering results (or the address data obtained as a result) become particularly important for establishing a ``probable cause'' to carry out subsequent searches and seizures at the suspect's premises \citep{deuber2022sok}.
The ``probable cause'' is a higher degree of suspicion and is prescribed by the US Constitution, 4th Amdt. to obtain a search or seizure warrant. Pursuant to this standard, law enforcement agencies must demonstrate that there is a connection between the requested evidence and the underlying offence, and that, based on the preliminary results of the investigation (factual basis), there are reasonable grounds to believe (probability forecast) that the requested evidence can be found at a specific location~\citep{christiansen2019forfeiting,dougherty2022united}. This requires a detailed presentation of the relevant evidence already available (e.g., a cryptocurrency analysis). The justification for this presumption of discoverability must be made on a case-by-case basis~\citep{moore1983fourth}, [Illinois v. Gates, 1983].``Investigative experience'' alone is not sufficient under US law or under German law.

\subsubsection{Second use case: Preliminary seizure of crypto assets} 

``Forfeiture'' (U.S.) or ``confiscation'' (Ger.) of assets is a procedural measure to deprive the suspect of the acquired goods. In both systems, this process is intended to ensure that those who commit crimes do not derive financial benefit from their actions~\citep{BrandKonstanz2023BGH,edgeworh2004asset,emehelu2018shot}, [Kaley v. United States, 2014]. 
The subcategories, ``civil forfeiture'' (U.S.) and ``extended confiscation'' (Ger.; GER-CC (German Criminal Code), Section 73a), enable the forfeiture/confiscation of additionally identified criminal assets in both systems (thus in different ways) ~\citep{serafin2021civil}. Their distinctive feature is that these procedural mechanisms also cover criminal assets that originally were not part of the investigation nor the crime being prosecuted. 
Unlike in German law, civil (judicial) forfeiture or ``non-conviction-based-forfeiture''~\citep{serafin2021civil} is only permitted in certain cases in the U.S.~\cite{christiansen2019forfeiting,cassella2019nature} (such as money laundering 18 U.S.C. Section 981 or child sexual abuse material 18 U.S.C. Section 2254). Another difference to German law is that civil forfeiture is not part of the criminal process but is pursued in a separate proceeding directed against the potentially incriminating assets themselves (action in rem) ~\citep{pimentel2012forfeitures,cassella2019nature,didwania2025asset,emehelu2018shot,christiansen2019forfeiting}. 

The reliability of the clustering method used becomes relevant in the context of the preliminary seizure of cryptocurrency assets, i.e., in the procedural stage that ensures the later judicial forfeiture/ confiscation: Once investigations have been narrowed to a single set of crypto addresses, investigators may observe subsequent transactions that involve additional suspicious (``tagged'') addresses. Under the reasonable assumption that these transferred crypto assets may also be linked to a previously unidentified criminal offence, they may be seized on a preliminary basis to enable a later ``civil forfeiture'' (U.S.) or ``extended confiscation'' (Ger.). Under U.S. law, the preliminary (civil) seizure of these assets underlies the low requirement that only the connection between the forfeitable~\citep{christiansen2019forfeiting} asset (located at the place to be searched) and a criminal offence must be established by the earlier described ``probable cause''~\citep{didwania2025asset,christiansen2019forfeiting,dougherty2022united}. While in Germany, prosecutors must raise the also already mentioned ``simple suspicion'' that these assets are linked to any criminal offence~\citep{Bittmann2023muncher111e,Meissner2025muncher} to execute the preliminary seizure in accordance with GER-CCP, Sections 111b et sqq. Because there are often no other indicators available at this stage, the required ``probable cause'' (U.S.)/``simple suspicion'' (Ger.) might be based solely on the clustering results. 

The preliminary seizure, therefore, marks another (but much more critical) use case in which the clustering results can even form the single basis for the required level of suspicion, and in which the reliability of the underlying clustering method must be considered by prosecutors to form the necessary degree of suspicion. 

\subsubsection{Third use case: Evidence in trial}

In an adversarial system, such as in the U.S., the parties are responsible for presenting all relevant evidence that supports their defence at court~\citep{eser2014adversatorische}. By contrast, in an inquisitorial system, such as in Germany, the judge bears responsibility for introducing, examining and evaluating all the evidence pertinent to the case~\citep{eser2014adversatorische}. 
Clustering results often play an indirect role in trial because they usually form the basis for earlier investigative measures, such as warrants or seizures, that lead to evidence relevant for trial proceedings.  

One of the rare examples where clustering results were used directly in trial, marks United States v. Sterlingov, 2024, in which the U.S. District Court for the District of Columbia accepted an expert testimony of a blockchain analysis company as evidence. In this case, to be admitted as evidence in court, clustering results were reviewed based on the reliability criteria of the ``Daubert Standard'' which was established through U.S. case law and is also partially included in the Federal Rules of Evidence, No. 702. It embodies an IT-Forensic Standard to measure the admissibility of digital evidence [Daubert v. Merrell Dow Pharmaceuticals, Inc., 1993]. According to this, parties are required to demonstrate in court whether an expert's reported method that provided the evidence can be tested, whether it has undergone peer review, what is known about any existing error rate and whether the method is widely accepted within a relevant scientific community (Daubert-Hearing) [Daubert v. Merrell Dow Pharmaceuticals, Inc., 1993]. Consequently, the reliability of the clustering method (s. ``known error rate'') is directly relevant for determining whether the parties meet the ``Daubert Standard'' and,  whether the presented clustering results will be accepted as evidence in court. Therefore, it must be proven by ``preponderance of the evidence'' that the clustering results are in fact based on reliable methods. This means that, according to the evidence presented by the parties (factual basis), it must more likely than not (probability forecast) that the clustering results are also the product of a reliable clustering method (s. also Federal Rules of Evidence. No. 702 (c)).

In Germany there is also an example where the evaluation of evidence in court was based primarily on the results of a blockchain analysis. (s. [Local Court of Aschaffenburg, judgment of November 09, 2023, case no. 307 Cs 115 Js 6602/23]). But, unlike U.S. law, German criminal procedure law is based on the principle of judicial free assessment of evidence rather than strict rules for measuring its admissibility (GER-CCP, Section 261)~\citep{BGHNJW19961420,BGHNJW19792311,BGHNJW19822882,ruckert2023digitale,deuber2022sok}. 
However, since judges must also base their evaluation of evidence on objective facts~\citep{BGHNJW19822882,BGHNJW20242856,BGHNSTZRR2025256}, commonly accepted scientific rules of experience, scientific experience based on scientifically proven probability values and other rules of experience become relevant in this context ~\citep{Bartel2024muncher,BGHNJW19792311,BGHNJW19571039,BGHNJW1967116,ruckert2023digitale}. Therefore, the reliability of the underlying clustering heuristic (such as the MIH) is also important for the judicial assessment under German law. 
If a reliability measure of a clustering heuristic could be determined, it would fall into the category of scientific experience based on scientifically proven probability values~\citep{ruckert2023digitale}. Depending on its quality, judges would then be bound by this probability value in court ~\citep{ruckert2023digitale}. However, if it is not possible to specify a general reliability measure for the MIH it would only represent other (non-certified) rules of experience~\citep{ruckert2023digitale}, which carry limited weight in the judicial assessment of evidence.

The reliability of the underlying clustering method may also influence the weight of the clustering results as evidence for the final judicial decision in trial.
If, after the Daubert-Hearing, (MIH-based) clustering results can be used as digital evidence in court in the U.S., the underlying clustering heuristic can also influence the final judicial decision as to whether the defendant committed the alleged offence ``beyond a reasonable doubt''. This judicial standard of proof is required for a criminal conviction and demands substantial and conclusive evidence (factual basis)  for each element of the alleged crime based on the evidence presented by the prosecution [In re Winship, 1970], i.e. that there is no logical explanation for the event other than the defendant's guilt (near certainty; very high probability forecast) [Patterson v. New York, 1977].
Similarly, under German law, a criminal conviction may only be handed down if the judge is convinced, based on the evidence considered at trial (factual basis), that the defendant committed the alleged offence (very high probability forecast) \citep{Bartel2024muncher}. 

\subsubsection{Resume and legal-technical proposals}

It was demonstrated that address clustering methods (such as the MIH) influence the factual basis and the probability forecast that underlie different degrees of suspicion and persuasion in criminal procedures. Therefore, from the perspective of prosecutors and judges who assess whether a sufficient factual basis and a sufficient probability forecast can be solely or additionally supported by clustering results, it is important to have an indication of the reliability of the underlying clustering method.

At least for the MIH, we attempt to determine the reliability of the underlying assumptions in practice. It has become apparent that, to date, a wide variety of metrics has been used to measure the performance of the MIH in scientific research. Additionally, we demonstrated that the performance of the MIH for the dataset average differs from its performance for individual entities. From a legal perspective, this is challenging because no general reliability measurement can be derived from these results.

We suggest that the legal and technical discussion should begin by agreeing on performance metrics that can serve as reliability benchmarks for address clustering methods in criminal proceedings. For the MIH, we recommend using ``per-wallet precision'' and ``per-wallet recall'' rather than relying on pairwise metrics, since real-world investigations often concern unlabeled addresses. We further suggest reporting both dataset-level averages and entity-level results to assess whether performance is homogeneous across entities or driven by strong variation. 
Per-wallet precision is particularly relevant because it indicates the extent to which the MIH produces false-positive matches, while per-wallet recall indicates the extent to which the MIH fails to recover true matches. 

Since legal probability forecasts for certain degrees of suspicion or persuasion cannot be translated into precise mathematical performance requirements, it is then up to prosecutors or judges to decide on a case-by-case basis whether a particular performance value for the MIH is sufficient or not to establish a certain level of suspicion or persuasion on MIH-based clustering results. 

However, as long as there is no consensus on this, the following best practices should be followed when applying address clustering methods (such as the MIH) in a legal context:
In addition to address clustering and investigations based on address-level, other methods should be employed to verify the clustering results. This approach has already been demonstrated in United States v. Sterlingov [United States v. Sterlingov, 2024], where different methods were used to verify the reliability of clustering results. But this also includes the use of additional clustering heuristics alongside MIH and additional information about the found addresses---if available. To perform a verification, the methodological details of the clustering heuristics used must first be disclosed.
Also, address clustering and investigations at the address-level should be conducted and interpreted in collaboration with qualified technical experts who are aware of the technical features of the blockchain and the clustering heuristics used. This involves the interpretation of clustering results as well as the analysis of the activities associated with the observed addresses.

\subsection{Limitations \& Future Work}

Our study has several limitations that open avenues for future work. First, the ground truth, while independent of heuristic assumptions, covers only seven services. One of our main results is that even these services differ, and other entities (e.g., in size, business model, or compliance obligations) might as well. Expanding the set of ground truth entities is an important next step. Only then should concrete recommendations for action be derived from these findings for law enforcement activities. Applying our framework to additional ground truth datasets would test whether the metric-dependent and entity-dependent patterns observed here generalize beyond this specific set of services.

Second, we evaluate a single, standalone heuristic (MIH), whereas real-world analytics platforms often combine multiple heuristics, machine learning models, and off-chain information. Our results thus provide a baseline for what MIH alone can achieve, not an end-to-end assessment of commercial tools. Future work should extend our results to evaluate combinations of heuristics and to study how additional information (e.g., service tags or temporal patterns) affects both performance and legal risk. The evaluation framework provided with this work can be used to assess the performance of other heuristics and tools on comparable ground truth datasets.

Third, our analysis is based on a static snapshot of the blockchain and ground truth. In practice, investigations unfold over time and may be affected by delays between on-chain activity and ground truth availability, as well as by changes in wallet software and user behavior. A standardized evaluation procedure should therefore be repeated regularly on updated ground truth datasets to verify whether address clustering heuristics and tools continue to perform reliably under changing blockchain usage patterns. The framework provided with this work can serve as a basis for such recurring evaluations by keeping the metric definitions and preprocessing steps comparable across time.

Finally, while the scope of this work is to reproduce and compare the metrics and reported numbers used in prior MIH evaluations, it is not to propose a new metric. However, our results indicate that existing metrics for the MIH only partially align with the requirements that law enforcement agencies face when justifying investigative measures based on heuristic outputs. An important direction for future work is therefore to agree on and consistently apply certain selected performance metrics in order to assess the reliability of MIH (and other address clustering methods).


\section{Conclusion}
\label{sec:conclusion}

The MIH is a foundational building block of Bitcoin address clustering and is widely used in research and in operational blockchain analytics. Yet, despite its central role and real-world consequences, MIH has only rarely been evaluated on ground truth, and existing evaluations are difficult to compare due to heterogeneous datasets and metrics. In this paper, we address this gap by reimplementing the preprocessing pipelines and nine evaluation metrics used in the three prior systematic MIH evaluations and applying them to a real-world ground truth dataset of seven services obtained independently of heuristic assumptions.

Our results show that MIH performance depends on the chosen metric and aggregation level. Dataset-level pairwise scores can appear strong while masking substantial failures for individual entities, and entity-level results reveal pronounced heterogeneity across services. We further demonstrate that pairwise metrics can be dominated by large services. Together, these findings caution against treating MIH-derived clusters as uniform or self-evident ground truth and highlight the need for evaluation practices that routinely report entity-level performance alongside global summaries.

From a law enforcement perspective, the costs of mainly false positives, as well as false negatives, make metric choice particularly consequential. To date, various calculation methods and reliability values for the MIH have been reported. However, there is currently no standardized consensus on which of these measurements are relevant for legal decision-making in the context of criminal proceedings.  An important direction for future work is therefore the development and standardization of a law enforcement-oriented evaluation procedure that can complement technical performance reporting and better support justified, accountable use of heuristic-based blockchain analytics.


\appendix



\begin{ethics}

This study uses publicly available Bitcoin blockchain data and pseudonymized ground truth data on address-to-entity mappings representing European crypto asset service providers. The ground truth data were obtained for research purposes and originate from legally mandated reporting obligations under the applicable regulatory framework. They were shared in pseudonymized form, with blockchain addresses and service-provider identifiers replaced by non-identifying labels.
Our institutions do not require IRB approval for this type of study, as it does not involve direct interaction with human subjects. Nevertheless, we considered the interests of affected stakeholders, including cryptoasset service providers, their customers, law enforcement agencies, and individuals who may be affected by blockchain forensic evidence. To reduce potential risks, we report results only in aggregate form and avoid any information that could reveal the identity of individual service providers or their customers.
The research aims to improve the reliability and transparency of blockchain forensic methods. We acknowledge that discussing limitations of the multi-input heuristic could inform malicious actors, but these limitations are already documented in prior work. We therefore consider the benefits for law enforcement, the judiciary, and affected individuals to outweigh this risk.

\end{ethics}

\begin{openscience}

The source code for our evaluation framework is available at \url{https://anonymous.4open.science/r/mih-toolkit}. The repository includes implementations of all nine metrics, preprocessing pipelines, scripts for extracting publicly available Bitcoin blockchain data, and synthetic examples illustrating the required input format.
The ground truth data used in this study consist of address-to-entity mappings obtained from European crypto asset service providers under legally mandated reporting obligations. Due to legal and confidentiality constraints, these data cannot be publicly released. They were provided exclusively for research purposes under a formal agreement that prohibits redistribution, and sharing them could reveal sensitive information about the involved service providers and their customers.
Although the ground truth data are restricted, researchers with access to comparable address-to-entity mappings can use our published code to replicate the evaluation methodology and report comparable results.


\end{openscience}

\begin{ai}
The authors used generative AI-based tools to revise the text, improve flow, and correct any typos, grammatical errors, and awkward phrasing. AI-based tools were used in the artifact development. We have manually verified and are responsible for the accuracy, originality, and integrity of the output of all AI-based tools.
\end{ai}

\bibliographystyle{ACM-Reference-Format}
\bibliography{references}







\end{document}